\documentclass[aps,prl,twocolumn,superscriptaddress,longbibliography]{revtex4-2}%
\usepackage{graphicx}
\usepackage{float}
\usepackage{dcolumn}
\usepackage{bm,color}
\usepackage{amsmath,amssymb,dsfont,amstext,amsfonts}
\usepackage{extarrows}
\usepackage{xcolor}
\usepackage{wasysym}
\usepackage{mathtools}
\usepackage{bbold}
\usepackage{hyperref}
\hypersetup{
breaklinks=true,
colorlinks=true,
citecolor=blue,
linkcolor=black,
filecolor=black,
urlcolor=blue
}
\usepackage[normalem]{ulem} 
\usepackage{color}

\def\tr{\mathrm{Tr}}

\newcommand{\mr}[1]{\mathrm{#1}}
\newcommand{\dg}{\dagger}
\newcommand{\mb}{\mathbf}

\begin{document}
\title{Gapped spinful phases obtained via Gutzwiller projections of Euler states}

\date{\today}

\author{Thorsten B. Wahl}
\email{tw344@cam.ac.uk}
\affiliation{TCM Group, Cavendish Laboratory, Department of Physics, J J Thomson Avenue, Cambridge CB3 0HE, United Kingdom}
\affiliation{Department of Physics and Astronomy, University of Manchester, Oxford Road, Manchester M13 9PL, United Kingdom}
\author{Lukas Devos}
\email{ldevos@flatironinstitute.org}
\affiliation{Flation Institute, 162 5th Ave., New York, NY 10010, U.S.A.}
\author{Robert-Jan Slager}
\email{rjs269@cam.ac.uk}
\affiliation{TCM Group, Cavendish Laboratory, Department of Physics, J J Thomson Avenue, Cambridge CB3 0HE, United Kingdom}
\affiliation{Department of Physics and Astronomy, University of Manchester,	Oxford Road, Manchester M13 9PL, United Kingdom}
\date{\today}

\begin{abstract}  
Gutzwiller projections of non-interacting chiral topological phases are known to give rise to fractional, topologically ordered chiral phases. 
Here, we carry out a similar construction using two copies of non-interacting Euler insulators to produce a class of spinful interacting Euler models. To that end, we take advantage of the recently discovered exact representation of certain Euler insulators by a projected entangled pair state (PEPS) of bond dimension $D = 2$. The Gutzwiller projection can be implemented within the tensor network formalism, giving rise to a new PEPS of bond dimension $D = 4$. We, moreover, apply very recent state-of-the-art tensor network tools to evaluate these phases. In particular, we analyze its entanglement entropy scaling and find no topological correction to the area law, indicating that the state is not intrinsically topologically ordered. Its entanglement spectrum shows a clear cusp at momentum zero, similar to non-interacting Euler insulators, and the spectrum of the transfer operator indicates that the state is gapped, which could imply non-intrinsic topological features. Finally, the static structure factor displays Bragg peaks, indicating the simultaneous presence of local order. 
\end{abstract}

\maketitle


\textit{Introduction. --} After the discovery of the quantum Hall effect in 1980 it became apparent that there are quantum phases of matter which fall beyond the Landau paradigm of local order parameters~\cite{vonKlitzing}. Almost 30 years later, insightful works~\cite{Schnyder2008,Kitaev} presented a complete classification of free fermionic topological phases~\cite{Rmp1,Rmp2} that include time reversal, chiral symmetry and particle hole symmetry as protecting symmetries based on K-theory. As a next step, the role of crystalline symmetries was considered, giving rise to a plethora of possible symmetry-protected free fermionic topological phases~\cite{Clas1,Clas2,Clas3,Clas4,Clas5,Clas6,PhysRevLett.117.096404, Shiozaki14}. Following these results, the past years have seen the rise of multi-gap topological phases that come about by virtue of having several bands involved~\cite{Clas7, bouhon2019nonabelian,BJY_nielsen, Ahn2019,Davoyan24, wu19} and can be classified by homotopy arguments~\cite{Clas7,Brouwer23}. Such multi-gap phases are interestingly related to non-Abelian charges defined in momentum space~\cite{wu19, BJY_nielsen, bouhon2019nonabelian}. That is, when considering more than two bands, adjacent pairs of bands can host band nodes that have non-Abelian charges that, after braiding them around in momentum space, can result in multi-gap invariants~\cite{BJY_nielsen, bouhon2019nonabelian}.

One of the most experimentally relevant such multi-gap invariants is the Euler class that arises due to combination of time reversal and reflection ($\mathcal{P} \mathcal{T}$ symmetry)~\cite{bouhon2019nonabelian, Clas7, Bouhon2018Wilson,BJY_nielsen, Ahn2019}. Euler insulators and semi-metals have been studied (and observed in trapped ions) in out-of-equilibrium~\cite{Unal_2020, Zhao_2022, Breach2024} and Floquet settings~\cite{slager2024floquet}, predicted in phonon spectra of materials such as silicates~\cite{Peng2021}, seen in acoustic meta-materials~\cite{Jiang_2021, JIANG2024,Guo1Dexp} and related to several electronic and magnetic materials and phases~\cite{bouhon2019nonabelian, BJY_nielsen, magnetic,lee2024eulerbandtopologyspinorbit, Konye21, chen22, ruegg2025eulertopologyclassicalspin}. Moreover, multi-gap phases are increasingly being related to novel responses by the aid of quantum geometrical interpretations~\cite{provost1980riemannian, bouhon2023quantum, jankowski2023optical,Jankowski24shift,avdoshkin2025multistategeometryshiftcurrent}. However, spinful interacting representatives of the above multi-gap topological phases or consequences of such interactions have so far remained elusive. 

The first interacting spinless multi-gap topological state was constructed in Ref.~\cite{EulerPEPS}, where an exact projected entangled pair (PEPS)~\cite{PEPS} representation of a non-interacting Euler insulator was found and subsequently made interacting via application of a shallow quantum circuit. This resulted in a new PEPS which is the unique ground state of an interacting Euler insulator in the same topological phase. A feature of this phase is a cusp in the entanglement spectrum at many-body momentum $K = 0$. Notably, all involved states were gapped and short-range correlated, as opposed to examples of chiral tensor network states~\cite{Wahl2013,Wahl2014,Dubail2015,Yang2015}. 

Here, we construct a representative class of spinful interacting gapped systems that arise by using the Euler PEPS as building blocks. Namely, we apply a Gutzwiller projection on two copies of a non-interacting Euler insulator that can be represented as a PEPS of bond dimension 2, as proposed in Ref.~\cite{EulerPEPS}. We use a Gutzwiller projection for two reasons: First, due to its locality, the new state is also an exact PEPS with bond dimension 4. Second, applying a Gutzwiller projection on two copies of a free fermionic topological state typically results in a state with spin degrees of freedom in a similar topological phase, potentially with topological order~\cite{Tu2013}. 

We calculate the entanglement spectrum of the new PEPS on the cylinder and find a cusp at many-body momentum $K = 0$, which is more clearly visible than the one of Ref.~\cite{EulerPEPS} for fermionic Euler insulators. We also compute the entanglement entropy as a function of the circumference of the cylinder and find no significant topological correction to the area law of entanglement. This indicates that the Gutzwiller projected state does not have topological order. We also calculate the gap of the transfer operator, which we find to be non-vanishing in the thermodyanmic limit. We thus conclude that the state we constructed has exponentially decaying correlations and is the ground state of a gapped, local Hamiltonian. These results suggest that the constructed PEPS has non-intrinsic topological features and might be a spinful representative of the non-interacting fermionic Euler phase. 
On top of that, we also calculate the structure factor, which displays pronounced Bragg peaks, indicating that the state also has local order. 


\textit{Construction. --} We first review the construction of free fermionic PEPSs that are unique ground states of Euler insulators~\cite{EulerPEPS} and thereafter describe how a Gutzwiller projection of two copies results in a new PEPS representing a spin system. 

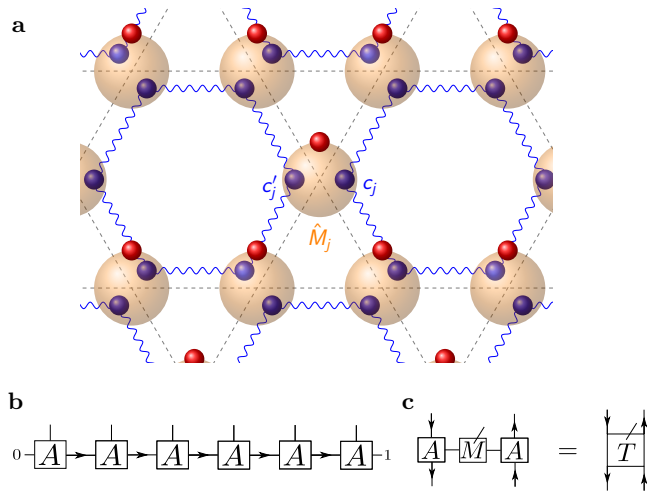
\begin{figure}[t]
	\begin{picture}(206,160)
		\put(10,30){\includegraphics[width=0.35\textwidth]{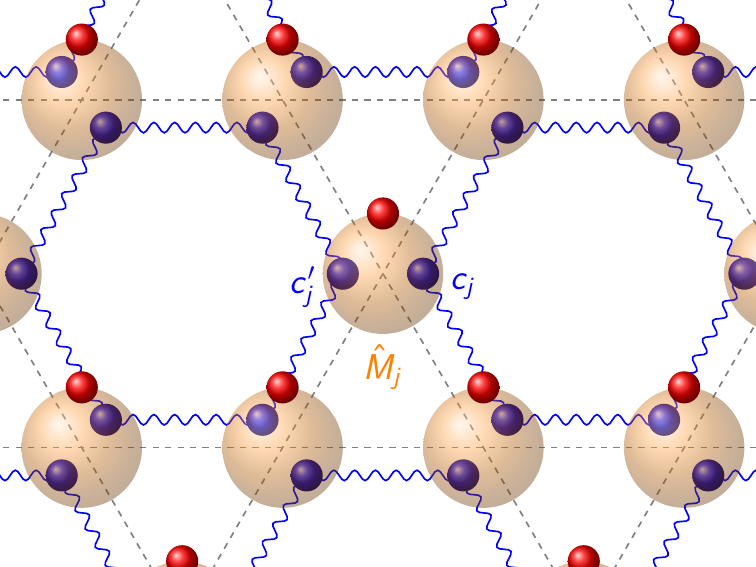}}
		\put(-16,155){\textbf{a}}
	\end{picture} 
	\begin{picture}(136,0)
		\put(-14,0){\includegraphics[width=0.28\textwidth]{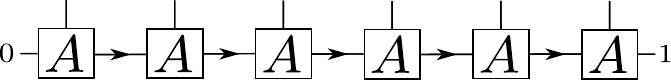}}
		\put(-16,23){\textbf{b}} 
	\end{picture} 
	\begin{picture}(70,0)
		\put(0,-8){\includegraphics[width=0.17\textwidth]{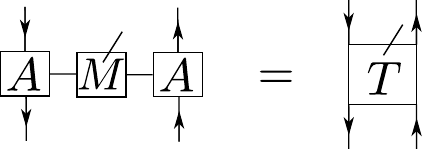}}
		\put(-6,23){\textbf{c}}
	\end{picture}
	\caption{a: Construction of the exact free fermionic Euler state~\cite{EulerPEPS}. The blue wiggly lines denote $W$-states of six virtual fermions ($c_j$). These get mapped to physical fermions (red balls) via $\hat M_j$ (shown as transparent orange spheres). b: Tensors representing $W$-states with $A_{01}^0 = 1/\sqrt{6}$, $A_{00}^1 = -A_{11}^1 = 1$ and all other entries equal to 0. The leftmost and rightmost virtual indices are fixed to 0 and 1, respectively. c: Overall tensor $T$ of the PEPS obtained by contracting two $A$-tensors with rank-3 tensor $M$, defined via $M_{11}^1 = M_{10}^0 = -M_{01}^0 = 1$ and all other entries equal to 0. All tensors have bond dimension $D = 2$. Arrows are only a visual aid to indicate which terminal virtual indices have to be set to $0$ and $1$, see Fig.~\ref{fig:PEPS}.}
	\label{fig:tensors}
\end{figure} 

We consider a system of spinless fermions on the kagome lattice of $N$ sites. The free fermionic PEPS we will consider can also be viewed as a projected entangled simplex state~\cite{PESS}: It is obtained from a virtual state with two fermions per site. Each of them is part of a $W$-state ($|011111\rangle + |101111\rangle + \ldots + |111110\rangle$)~\cite{W-state} of virtual fermions that lie on hexagons, see Fig.~\ref{fig:tensors}a. Finally, the two virtual fermions per site are mapped to one physical fermion via the map $\hat M_j = a_j^\dg c_j' c_j + c_j - c_j'$. Here, $c_j^{(\prime)}$ corresponds to the virtual fermion right (left) of site $j$ and $a_j$ to the physical one. Finally, the virtual fermions are projected on their vacuum state $|0_v \rangle$,
\begin{align}
	|\psi_\mr{ferm} \rangle = \langle 0_v| \prod_{j=1}^N \hat M_j \prod_{\hexagon} \frac{1}{\sqrt 6} \sum_{i=1}^6 c_{\hexagon,i} |1_v 0_p\rangle, \label{eq:fPEPS}
\end{align}
where the index $p$ ($v$) refers to all physical (virtual) fermions and $i = 1, \ldots, 6$ to the index within hexagon $\hexagon$. The state $|\psi_\mr{ferm}\rangle$ has two occupied bands with non-trivial Euler topology~\cite{EulerPEPS}. The PEPS is the unique ground state of a free fermionic three-band Hamiltonian if the chemical potential lies in the gap between the lowest two and the top third band. The corresponding PEPS tensor $T$ for each site of the kagome lattice is obtained by contracting two tensors $A$ representing $W$-states with a tensor $M$ representing the map $\hat M_j$, see Fig.~\ref{fig:tensors}b,c. Finally, the spinful PEPS is obtained by applying a Gutzwiller projection on two copies of \eqref{eq:fPEPS} 
\begin{align}
	|\psi_\mr{spin}\rangle = P_G |\psi_{\mr{ferm}}^\uparrow\rangle |\psi_{\mr{ferm}}^\downarrow \rangle, \label{eq:sPEPS}
\end{align}
with $P_G = \prod_j \left( a_{j \uparrow}^\dg a_{j\uparrow} a_{j \downarrow}^\dg a_{j \downarrow} +  a_{j \uparrow} a_{j\uparrow}^\dg a_{j \downarrow} a_{j \downarrow}^\dg \right)$ the Gutzwiller projector, where $\uparrow, \downarrow$ label the two copies. It enforces that each site is either double occupied or empty, effectively turning the system into a spin-$1/2$ model. We note that the resulting state is related to the one proposed in Ref.~\cite{EulerPEPS} (using a Gutzwiller projection that enforces single occupancies and particle-hole transformation) via an on-site unitary transformation. We also note that the construction presented here can be easily generalized to other states by using $\sum_{i=1}^6 \beta_i c_{\hexagon,i}|1_v\rangle$ with $\beta_{i+3}^* = \beta_i \in \mathbb{C}$ as the virtual simplex state in Eq.~\eqref{eq:fPEPS}. 

\begin{figure}[t]
	\begin{picture}(200,50)
		\put(50,0){\includegraphics[width=0.25\textwidth]{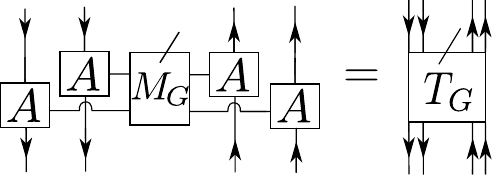}}
		\put(-20,45){\textbf{a}}
	\end{picture} 
	\begin{picture}(200,130)
		\put(-20,0){\includegraphics[width=0.48\textwidth]{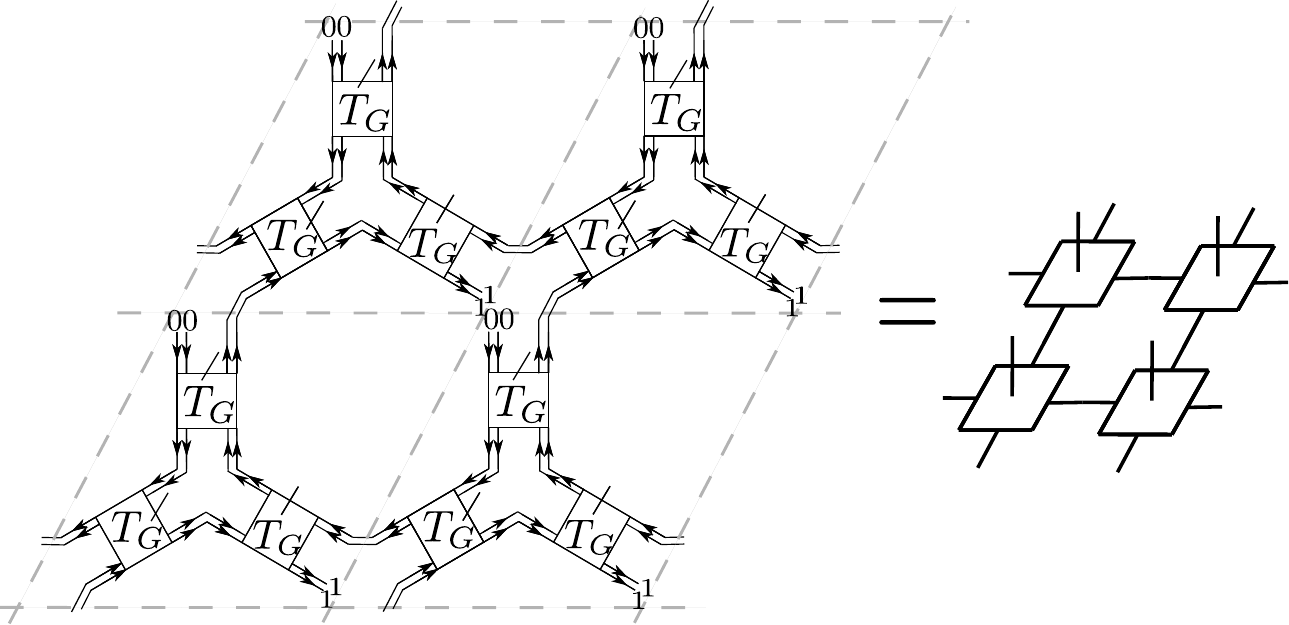}}
		\put(-20,115){\textbf{b}} 
	\end{picture} 
	\caption{a: Construction of the PEPS tensor $T_G$ of $|\psi_\mr{spin}\rangle$. b: Obtaining a square lattice PEPS. As some of the virtual indices have to be set to $0$ and $1$, blocking all three tensors of a unit cell (shown by dashed lines) results in a PEPS tensor defined on a square lattice with three physical spin-$1/2$ degrees of freedom and still bond dimension $D = 4$.}
	\label{fig:PEPS}
\end{figure} 

The probably easiest way to calculate the PEPS tensor corresponding to $|\psi_\mr{spin}\rangle$ is first to apply the Gutzwiller projector $P_G$ on two copies of $\hat M_j$, 
\begin{align}
\hat M_{G,j} &\equiv P_G \hat M_j \hat M_j \notag \\
&= a_{j,\uparrow}^\dg a_{j,\downarrow}^\dg c_j' c_j d_j' d_j + (c_j - c_j')(d_j - d_j'),
\end{align} 
where $d_j$, $d_j'$ are the virtual fermions of the second copy. 
The new PEPS tensor, say $T_G$, results from the contraction of the tensor corresponding to $\hat M_{G,j}$ with two copies of $A$ on the left and two on the right, see Fig.~\ref{fig:PEPS}a. Finally, there is a very convenient way of transforming the PEPS, which is defined on a kagome lattice, to one defined on a square lattice by blocking all three tensors $T_G$ of a unit cell together and taking advantage of the boundary conditions indicated in Fig.~\ref{fig:tensors}b. This is shown in Fig.~\ref{fig:PEPS}b. The representation of Eq.~\eqref{eq:sPEPS} as a standard square lattice PEPS allows us to use existing fermionic PEPS codes to analyze its features. (We note that due to the virtual fermions, the PEPS construction is still fermionic in nature.) For the calculations on a cylinder below, we used an extended version of the fermionic PEPS code presented in Ref.~\cite{Yang2015} and for the calculations in the infinite plane we used TensorKit.jl and PEPSKit.jl~\cite{Haegeman2025,Brehmer2025}.


\textit{Results. --} Here, we present our analysis of the topological correction to the area law of entanglement, the entanglement spectrum, the gap of the transfer operator, and the static structure factor of $|\psi_\mr{spin}\rangle$. 

We first consider a finite system with periodic boundary conditions, i.e., a torus geometry. For the entanglement entropy and spectrum, the torus is cut in half and we are interested in the spectrum of the reduced density matrix corresponding to one half. The entanglement spectrum can be obtained by tracing out the physical spins and placing the PEPS on a (horizontal) cylinder, using a simple virtual state as an input state on the left (e.g. the vacuum) and contracting from the left to the right. This results in a reduced density matrix $\sigma_R$. The same procedure contracting from the right to the left results in a reduced density matrix $\sigma_L$. For sufficiently many rings, $\sigma_{L}$ and $\sigma_R$ are converged. The non-trivial spectrum of the physical reduced density matrix is given by~\cite{Cirac2011} $\sqrt{\sigma_L^\top} \sigma_R \sqrt{\sigma_L^\top} \propto \rho_L := e^{-H_L}$ with $\tr(\rho_L) := 1$. $H_L$ denotes the entanglement Hamiltonian encoding the entanglement spectrum, and the entanglement entropy is obtained via $S = - \tr[\rho_L \ln (\rho_L)]$. For details, see Ref.~\cite{Yang2015}.  

\begin{figure}[t]
	\includegraphics[width=0.5\textwidth]{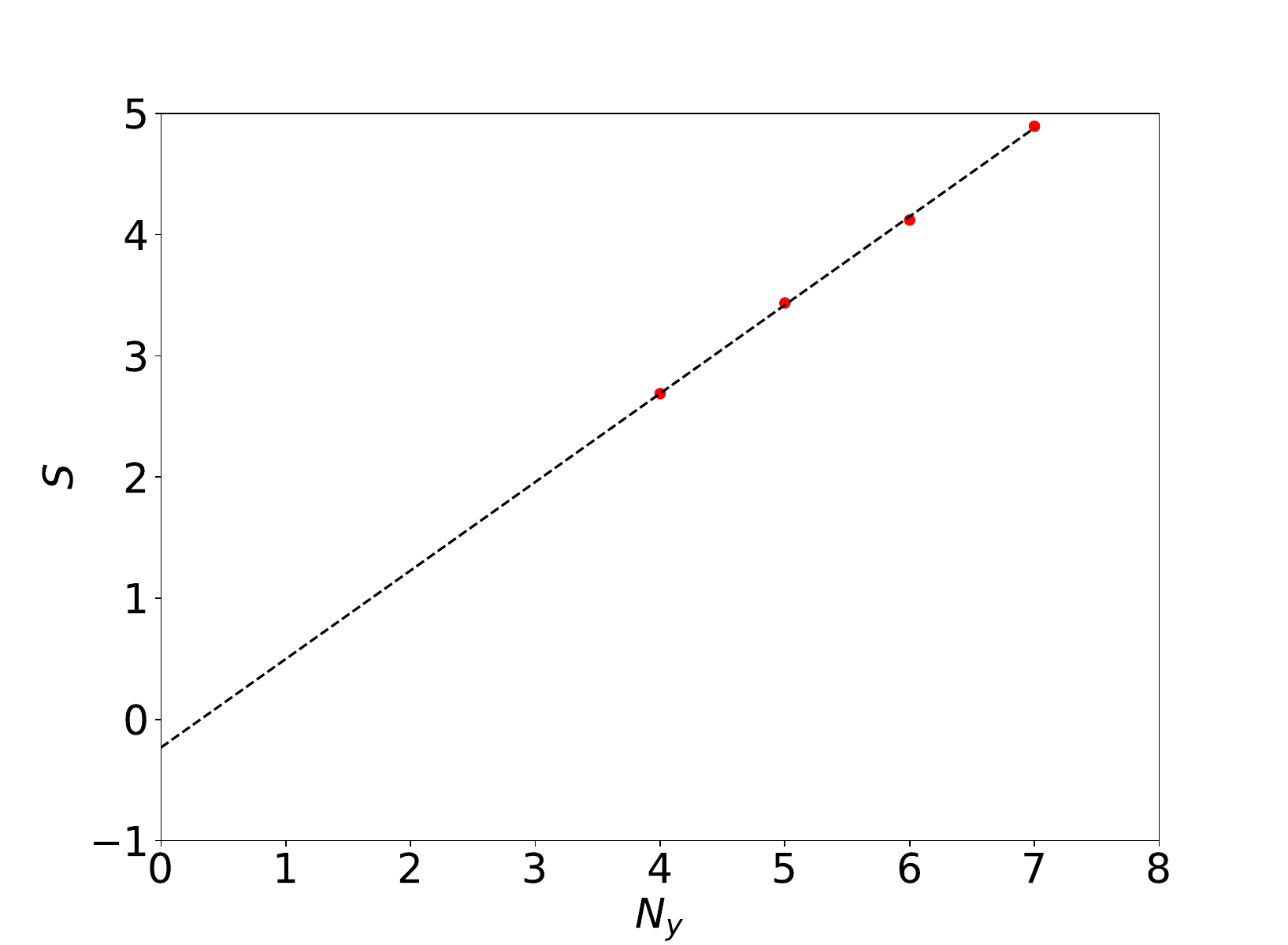}
	\caption{Entanglement entropy as a function of cylinder circumference (red dots). The dashed line indicates a linear fit. It has an intercept of $-0.23$.}
	\label{fig:area_law}
\end{figure}

\begin{figure}[t]
	\includegraphics[width=0.5\textwidth]{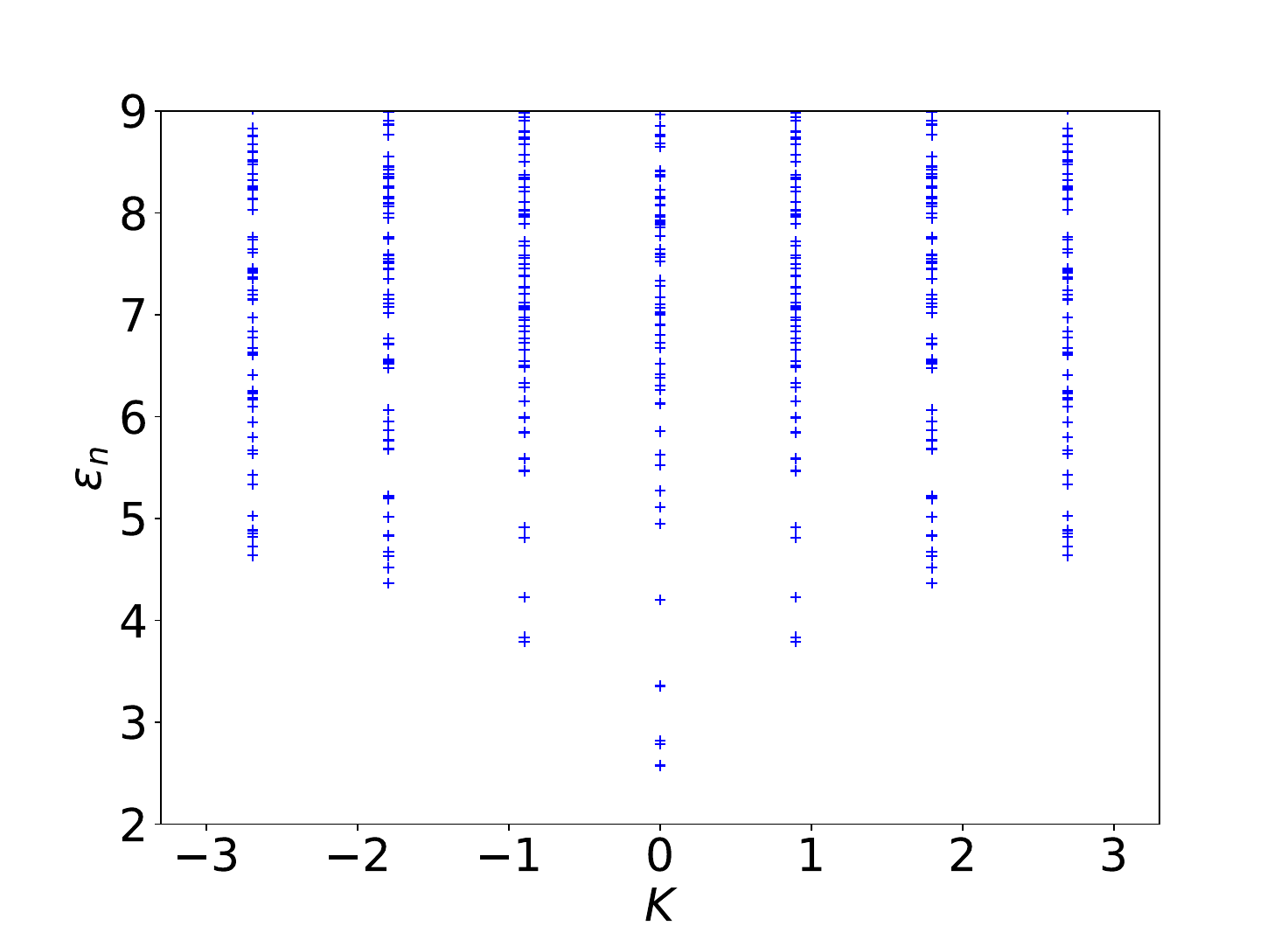}
	\caption{Entanglement spectrum for a cylinder circumference $N_y = 7$. $K$ denotes the many-body momentum and $\epsilon_n$ the entanglement energies (eigenvalues of $H_L$). The spectrum was well converged for a cylinder of length $30$.}
	\label{fig:entspec}
\end{figure}

In Fig.~\ref{fig:area_law}, we present our results on the entanglement entropy as a function of cylinder circumference $N_y$ (number of unit cells). A linear fit indicates a correction to the area law of entanglement $S(\rho_L) = c N_y - \gamma$ of $\gamma = 0.23$. This is inconsistent with a topologically ordered state, which has $\gamma = \ln(\mathcal{D})$, where $\mathcal{D} = 2,3,\ldots$ is the quantum dimension and ground state degeneracy on the torus. The small, non-zero correction we obtain thus has to be attributed to finite-size effects. We conclude that our state is not topologically ordered. Nevertheless, its entanglement spectrum has similarities to the one of a \mbox{(non-)}interacting fermionic Euler system reported in Ref.~\cite{EulerPEPS}: Specifically, we obtain a gapless spectrum with a cusp at many-body momentum $K = 0$, cf. Fig.~\ref{fig:entspec}. The entanglement spectrum of the interacting Euler PEPS analyzed in Ref.~\cite{EulerPEPS} shows similar characteristics, with an isolated entanglement energy at $K = 0$ separated by a small gap from the dense entanglement spectrum above. In that reference, some of us conjectured that for larger cylinder circumferences, the $K = 0$ mode will be connected to the spectrum above, forming a cusp, as is observed in this work. Based on the results of Ref.~\cite{EulerPEPS}, this cusp is a signature of the Euler phase, even in the presence of interactions. As we see the same signature in the present work, but no topological order, the state we constructed is presumably a spin representative of the non-interacting fermionic Euler phase. 

\begin{figure}[t!]
	\begin{picture}(200,171)
		\put(-3,0){\includegraphics[width=0.44\textwidth]{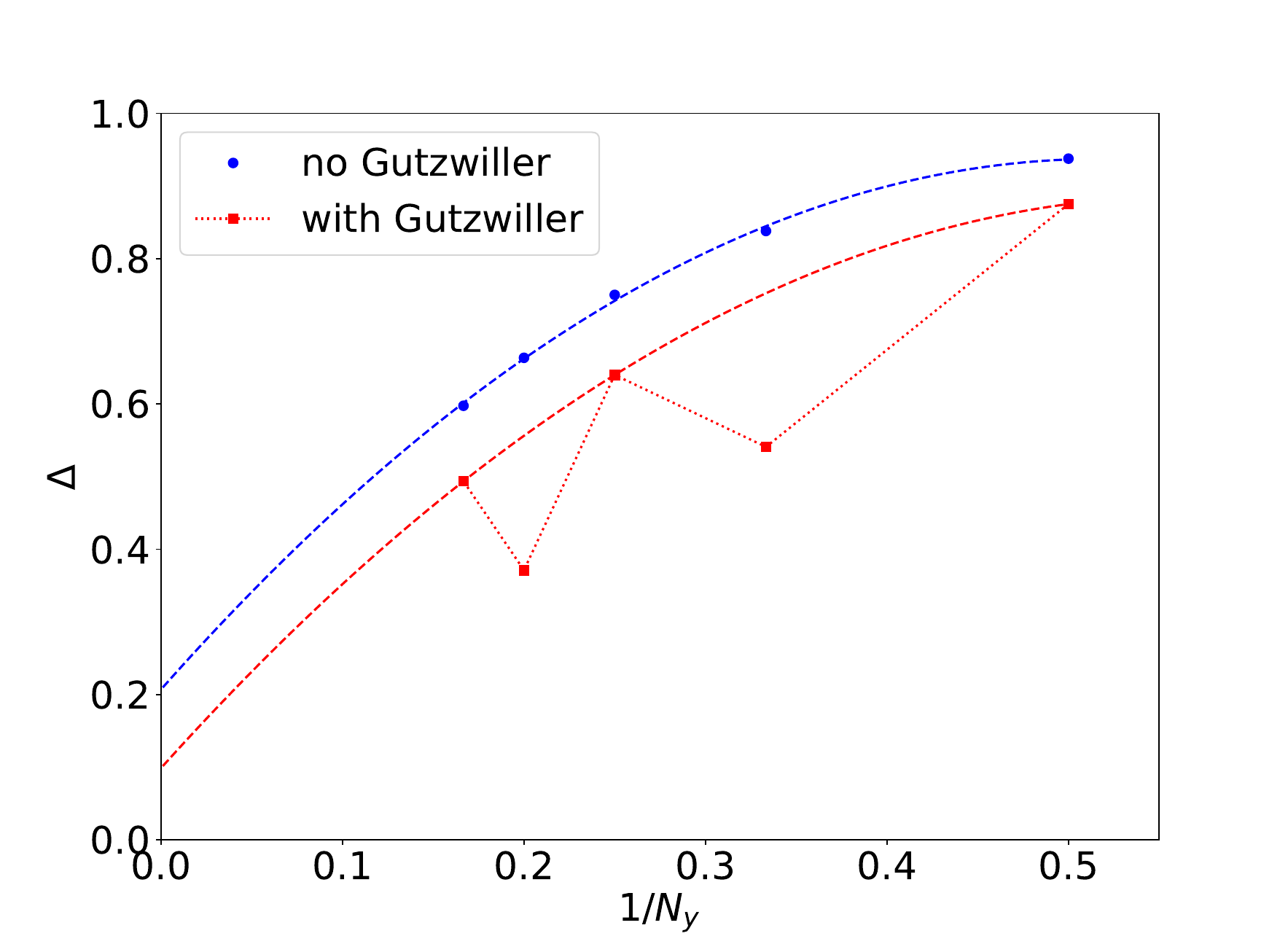}}
    \put(-2,155){\textbf{a}}
	\end{picture} \\
    \begin{picture}(200,155)
		\put(0,0){\includegraphics[width=0.43\textwidth]{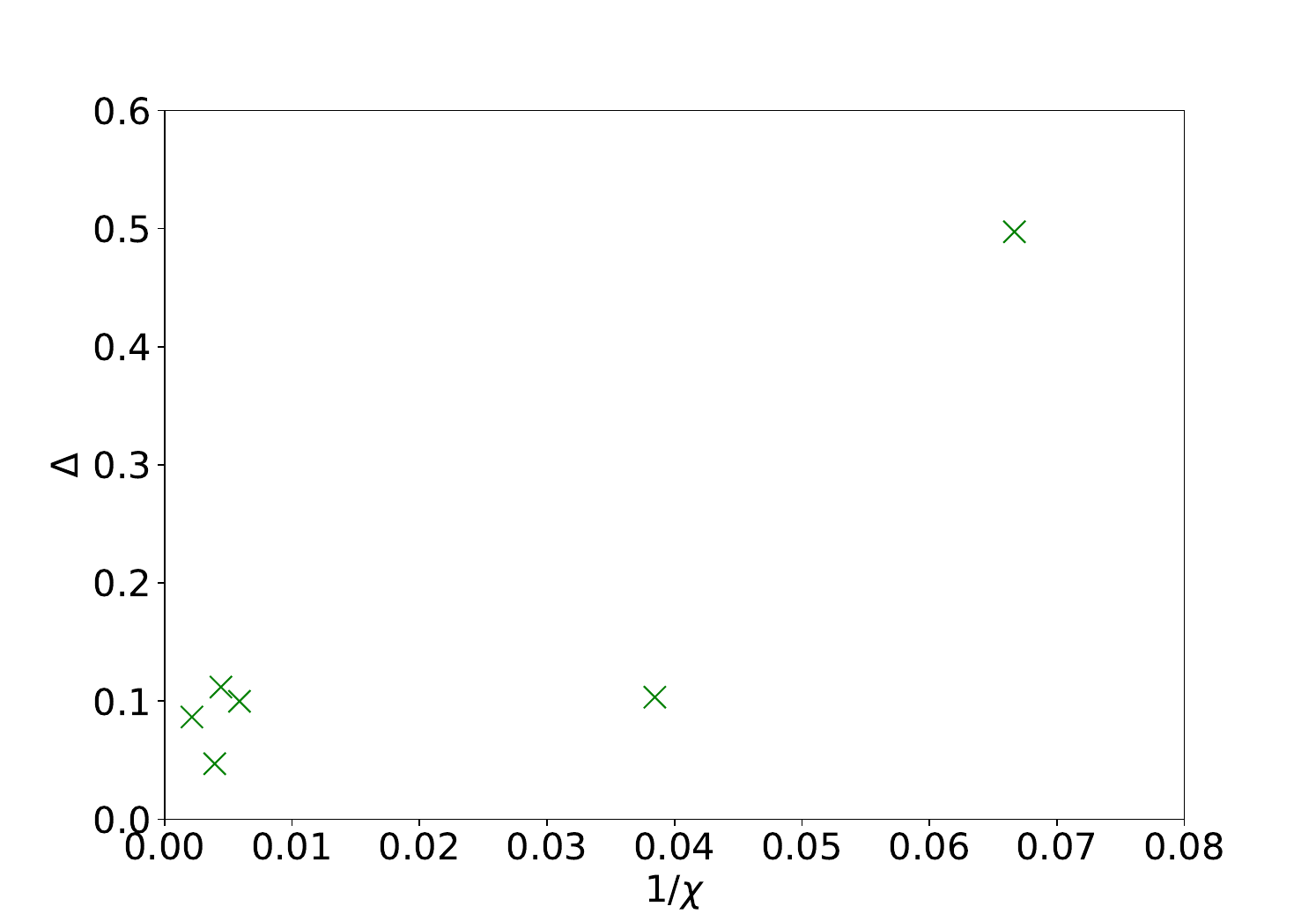}}
    \put(-2,146){\textbf{b}} 
    \end{picture}
	\caption{a: Gap of the transfer operator as a function of the inverse cylinder circumference $N_y$. For comparison, we show the same results with (red) and without (blue) the application of the Gutzwiller projector. 
    The dashed lines indicate inverse quadratic fits ($a_0 N_y^{-2} + a_1 N_y^{-1} + a_2$) to extrapolate to the thermodynamic limit (only using even $N_y$ for the Gutzwiller projection). While we indeed find a small gap for $N_y \rightarrow \infty$ without Gutzwiller projection, even-odd fluctuations prevent clear conclusions in the Gutzwiller-projected case. 
    b: Gap of the row-to-row transfer operator on the infinite plane as a function of the inverse CTMRG environment dimension $\chi$, indicating a finite gap in the limit $\chi \rightarrow \infty$.} \label{fig:transfer_gap}
\end{figure}

One important question is whether the investigated state has exponentially decaying correlations, i.e., corresponds to the ground state of a gapped Hamiltonian. To that end, we calculated the gap of the transfer operator on the cylinder, which is conjectured to reflect the presence of a gap in the spectrum of a parent Hamiltonian for the PEPS. This amounts to calculating the eigenvalues of the application of a single ring of the cylinder (with physical spins traced out) on a virtual input state $\sigma$. The resulting gap between the leading ($=1$) and second highest eigenvalue can be gathered from Fig.~\ref{fig:transfer_gap}a. For comparison, we also show the result for $|\psi_{\mr{ferm}}\rangle$ in the same figure, which displays a small gap in the extrapolated limit $N_y \rightarrow \infty$. However, sizable even-odd fluctuations for $|\psi_{\mr{spin}} \rangle$ prevent drawing clear conclusions as to the presence of a gap in the thermodynamic limit. 

We thus supplement our study with a calculation directly in the thermodynamic limit, making use of the PEPS tensor on the square lattice as shown in Fig.~\ref{fig:PEPS}b. The required $\mathbb{Z}_2$-graded representation~\cite{Mortier2025} of the tensor is derived in the Supplementary Information (SI)~\cite{SI}. Here, we first approximate the appropriate transfer operator through a corner transfer matrix renormalization group (CTMRG)~\cite{Nishino1996,Fishman2018} computation, for which we can tune the accuracy through the environment dimension $\chi$. We can then obtain the value of the gap in a similar fashion as on the cylinder. The results are presented in Fig.~\ref{fig:transfer_gap}b, indicating a gap in the limit $\chi \rightarrow \infty$, and we elaborate on these methods in the SI~\cite{SI}. Combining the results on the infinite plane and the cylinder, we conclude that our findings are consistent with the existence of a gap in the thermodynamic limit.

\begin{figure}[t]
	\includegraphics[width=0.4\textwidth]{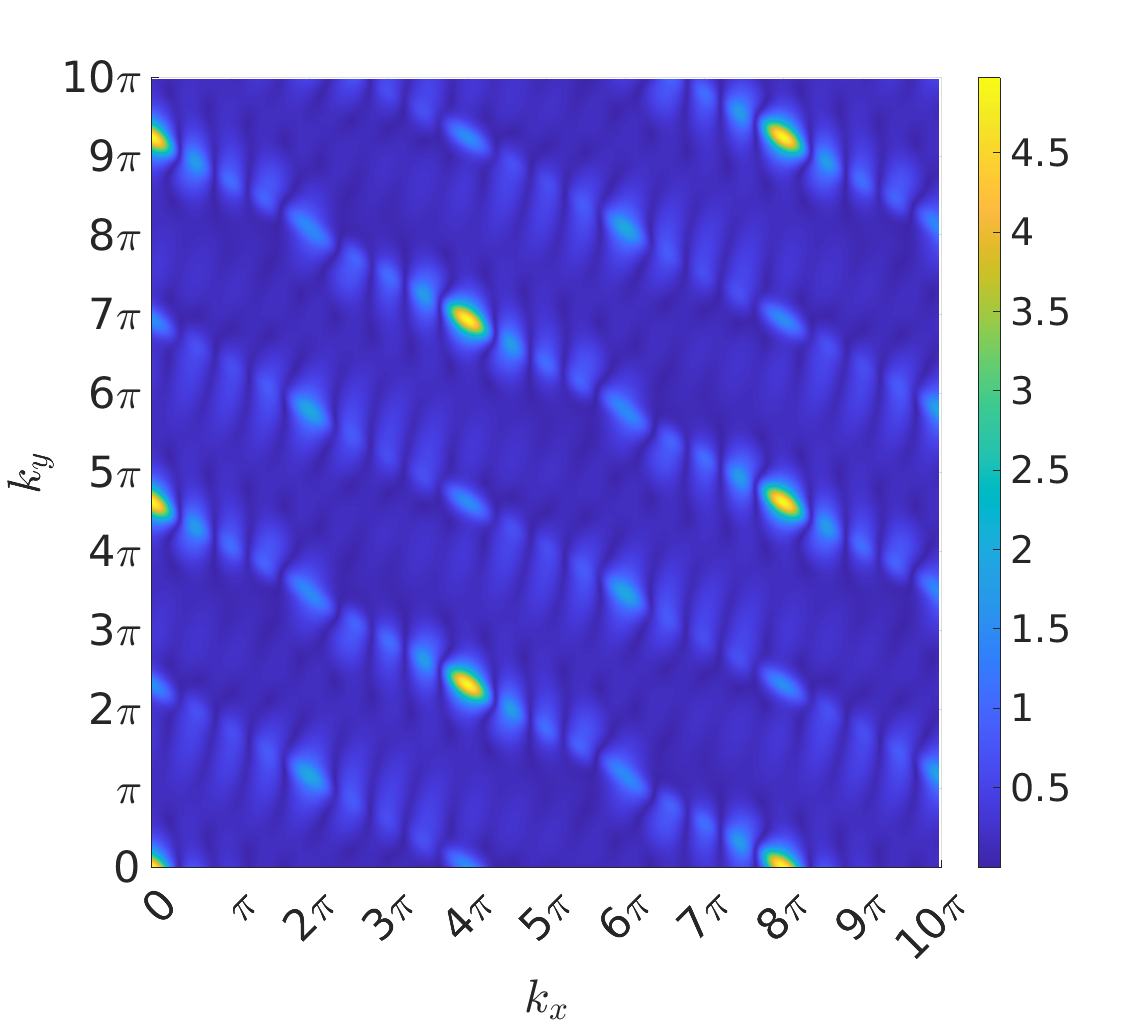}
	\caption{Magnitude of the structure factor $|S(\mb k)|$ obtained by contracting the PEPS on a cylinder of circumference $N_y = 5$ and length 41. Bragg peaks are clearly visible, indicating the presence of local order.}
    \label{fig:structure_factor}
\end{figure}

Finally, to check for the presence of local order in the state $|\psi_\mr{spin}\rangle$, we calculated the static structure factor~\cite{Zheng2005}, defined as
\begin{align}
S(\mb k) = \sum_{\mb r, \alpha} e^{i \mb k \cdot \mb r} \left( \langle S_\mb{0}^\alpha S_\mb{r}^\alpha \rangle - \langle S_\mb{0}^\alpha \rangle \langle S_\mb{r}^\alpha \rangle\right),
\end{align}
where $S^\alpha_\mb{r} = \frac{1}{2}\sigma^\alpha_\mb{r}$, $\alpha = x,y,z$, is the spin operator. 
Specifically, we compute the corresponding expectation values by placing the PEPS on a cylinder of circumference $N_y = 5$ and of length 41. We put the central site $\mb 0$ in the middle of the cylinder and numerically sum over all sites $\mb r$ in the same ring as site $\mb 0$ and ten further rings in both directions. The correlations turn out to decay quickly with distance. The result is shown in Fig.~\ref{fig:structure_factor} with clear Bragg peaks. We conclude that our state also possesses local order. 

\textit{Conclusions. --} We investigated the effect of applying a Gutzwiller projection on two copies of a non-interacting Euler insulator that can be exactly represented as a PEPS. Using the tensor network formalism, we studied the entanglement entropy, entanglement spectrum, transfer operator gap, and structure factor of the resulting state. 
We found evidence that our state has an energy gap and displays similar topological features as the non-interacting fermionic Euler phase. Consistent with that, the entanglement entropy scaling indicated the absence of topological order. These features make it likely that our PEPS is a symmetry-protected topological state, specifically, a spin representative of the non-interacting fermionic Euler phase. Furthermore, the structure factor shows Bragg peaks, demonstrating that the state also has local order. It therefore remains to be investigated if there are topologically ordered Euler phases. 
We leave this question for future work.

\begin{acknowledgements}
This project was supported by funding from an EPSRC ERC underwrite grant EP/X025829/1 (T.B.W. and R.-J. S.) and a Royal Society exchange grant IES/R1/221060.
The Flatiron Institute is a division of the Simons Foundation (L.D.).
\end{acknowledgements}

\bibliography{references}{}

\begin{appendix}

\begin{widetext}
\section{Explicit form of the PEPS tensor} 

Here, we present the explicit form of the tensors in $\mathbb{Z}_2$-graded representation~\cite{Mortier2025}. The resulting overall PEPS tensor was used in the calculations with TensorKit~\cite{Haegeman2025}.

We first give the $\mathbb{Z}_2$-graded form of the $A$-tensors representing the $W$-states. One can check that the following tensor gives rise to the right result for a matrix product state (MPS)~\cite{MPS} defined on six sites
\begin{align}
A = |1,0\rangle \langle 0| - |1,1 \rangle \langle 1| + \frac{1}{\sqrt{6}}|0,0\rangle \langle 1|,
\end{align}
where the ket $|n_c,l\rangle$ corresponds to an output state in $|n_c\rangle$ ($n_c$ is the number of fermions of type $c$) and left input state $\langle l |$. Similarly, the bra $\langle r|$ corresponds to the right input state $|r\rangle$ of the MPS. We can easily check
\begin{align}
\langle 0 | A A |1\rangle &= \frac{1}{\sqrt{6}}\left(|10\rangle + |0 1\rangle \right), \notag \\
\langle 0 | A A A |1 \rangle &= \frac{1}{\sqrt{6}}\left(|110\rangle + |101\rangle + |011\rangle\right),  \\
&\ldots \notag
\end{align}
The map from the virtual to the physical space (before Gutzwiller projection) can be written as
\begin{align}
M = |1\rangle\langle 11| + |0 \rangle \langle 1 0| - |0\rangle \langle 0 1|,
\end{align}
where we have the ordering $|n_a\rangle \langle n_{c'} n_c|$ with $n_a$ corresponding to the number of physical $a$-fermions and $n_{c^{(\prime)}}$ corresponding to the number of virtual $c^{(\prime)}$-fermions. As we will use two copies of the free fermionic Euler PEPS, we construct the graded tensor product of the tensor $A$ with itself,
\begin{align}
    A_2 := A \otimes_g A &= \left(|1,0\rangle \langle 0| - |1,1 \rangle \langle 1| + \frac{1}{\sqrt{6}} |0,0\rangle \langle 1|\right) \left(|1,0\rangle \langle 0| - |1,1 \rangle \langle 1| + \frac{1}{\sqrt{6}} |0,0\rangle \langle 1|\right) \notag \\
    &= |11,00\rangle \langle00| - |11,01\rangle \langle 01| + \frac{1}{\sqrt{6}}|10,00\rangle\langle 01| - |11,10\rangle\langle 10| - |11,11\rangle \langle 11| - \frac{1}{\sqrt{6}}|10,10\rangle \langle 11| \notag \\
    &- \frac{1}{\sqrt{6}} |01,00\rangle \langle 10| - \frac{1}{\sqrt{6}}|01,01\rangle \langle 11| + \frac{1}{6}|00,00\rangle \langle 11|,
\end{align}
where the kets and bras are ordered $|n_c n_d, l L\rangle \langle r R|$ with $n_c,l,r$ ($n_d,L,R$) corresponding to the occupations of the first (second) copy. Similarly, for the graded tensor product of the map $M$ with itself, we have
\begin{align}
    M_2 := M \otimes_g M &= \left(|1\rangle\langle 11| + |0 \rangle \langle 1 0| - |0\rangle \langle 0 1|\right) \otimes_g \left(|1\rangle\langle 11| + |0 \rangle \langle 1 0| - |0\rangle \langle 0 1|\right) \notag \\
    &= - |11\rangle \langle 1111| - |10\rangle\langle 1110| - |10\rangle \langle 1011| - |01\rangle \langle 1101| + |00\rangle \langle 1100| - |00\rangle \langle 1001| - |01\rangle \langle 0111| \notag \\
    &+ |00\rangle \langle 0110| + |00\rangle \langle 0011|,
\end{align}
where the kets and bras are ordered $|n_{a_\uparrow} n_{a_\downarrow} \rangle \langle c' d' c \, d|$ with $a_{\uparrow},c,c'$ ($a_\downarrow,d,d'$) corresponding to the fermions of the first (second) copy. The Gutzwiller projector enforces that each site is either empty or doubly occupied, i.e.,
\begin{align}
M_G := P_G M_2 = -|11\rangle \langle 1111| + |00\rangle \langle 1100| - |00\rangle \langle 1001| + |00\rangle \langle 0110| + |00\rangle \langle 0011|.
\end{align}
Finally, the PEPS tensor $T_G$ (see Fig.~\ref{fig:PEPS}b) corresponding to one site is obtained as
\begin{align}
T_G = M_G (A_2 \otimes_g A_2),
\end{align}
where the graded tensor product now refers to a product between the pair of $A$-tensors left of $M_G$ and the pair right of $M_G$ in Fig.~\ref{fig:PEPS}a (i.e., between the vector spaces of the unprimed and primed fermions $c,l,r,d,L,R$). We do not give the expression for $T_G$ explicitly here, but we know that it is of the form
\begin{align}
T_G = \sum_{n,l',L',l,L,r',R',r,R}  c_{nl'L'lLr'R'rR} |n, l' L' l L \rangle \langle r' R' r R |
\end{align}
with $n = n_{a_\uparrow} = n_{a_\downarrow}$. The (square lattice) tensor corresponding to one unit cell is 
\begin{align}
T_{3G} &= \langle l_1' = 0, L_1' = 0| \left( \sum_{\{n_i,l_i^{(\prime)},L_i^{(\prime)},r_i^{(\prime)},R_i^{(\prime)}\}}  c_{n_1l_1'L_1'l_1L_1r_1'R_1'r_1R_1} |n_1, l_1' L_1' l_1 L_1 \rangle \langle r_1' R_1' r_1 R_1 | \times \right. \notag \\
& c_{n_2r_1'R_1'l_2L_2r_2'R_2'r_2R_2} |n_2, r_1' R_1' l_2 L_2 \rangle \langle r_2' R_2' r_2 R_2 | \ c_{n_3r_2 R_2l_3L_3r_3'R_3'l_1L_1} |n_3, r_2 R_2 l_3 L_3 \rangle \langle r_3' R_3' l_1 L_1 | \Biggr) | r_3' =1,R_3' = 1\rangle,
\end{align}
see Fig.~\ref{fig:T3G}.

\begin{figure}[t]
	\includegraphics[width=0.6\textwidth]{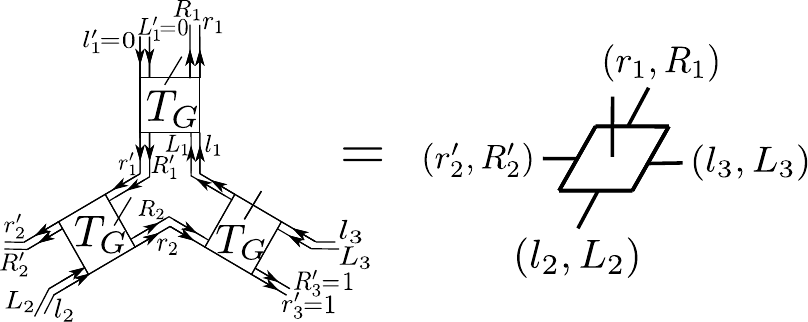}
	\caption{Blocking of the three PEPS tensors in a unit cell $T_G$ to obtain the overall PEPS tensor of the square lattice $T_{3G}$.}
	\label{fig:T3G}
\end{figure}

\section{Calculations on the infinite plane} 

In order to show that the constructed PEPS results in exponentially decaying correlations in the thermodynamic limit, we combine standard methods for infinite PEPS with the formalism outlined in \cite{Mortier2025} to include the fermionic symmetry of the tensors.
This is implemented in the open-source library PEPSKit.jl \cite{Brehmer2025}, which is built on top of TensorKit.jl \cite{Haegeman2025}.
In particular, without loss of generality, we may restrict ourselves to evaluating horizontal and vertical two-point functions $\langle O^{(1)}_{0,0}, O^{(2)}_{n,0}\rangle$ and $\langle O^{(1)}_{0,0}, O^{(2)}_{0,n}\rangle$ for arbitrary operators $O^{(1)}$ and $O^{(2)}$.
If these show exponential decay, also any diagonal two-point function will be exponentially suppressed.
Focusing on the horizontal case, in order to evaluate such expressions, we first combine the double-layer network of tensors ${T_{3G}^*}$ and $T_{3G}$ into a single-layer tensor $A_\mr{peps}$ to simplify the following discussion (noting that our implementation does keep the layers separate to ensure maximal efficiency).
We may then reduce the contraction on the infinite plane to a quasi-one-dimensional structure using the four CTMRG environment corner and edge tensors $C_i$ and $E_i$, one for each direction.
Then, we can construct partial contractions $L(O^{(1)})$ and $R(O^{(2)})$ and the horizontal transfer operator $T_H(E_{\text{north}}, A_{\text{peps}}, E_{\text{south}})$ to evaluate the two-point function as
\begin{equation}
    \label{eq:peps_twopoint}
    \langle O^{(1)}_{0,0}, O^{(2)}_{0,n} \rangle = L(O^{(1)}) \left(T_H(E_{\text{north}}, A_{\text{peps}}, E_{\text{south}})\right)^n R(O^{(2)}).
\end{equation}
To clarify, the CTMRG environment tensors and the definition of the partial contractions are illustrated in Fig.~\ref{fig:ctmrgtensors}.

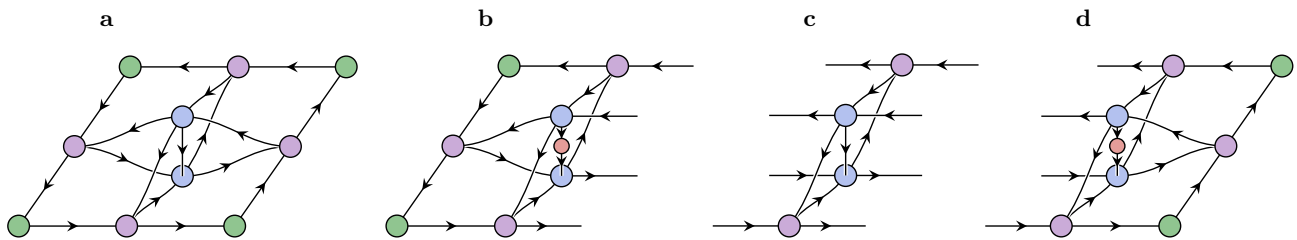
\begin{figure}
	\begin{picture}(140,100)
		\put(0,0){\includegraphics[width=0.26\textwidth]{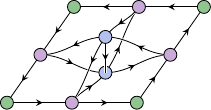}}
    \put(35,80){\textbf{a}}
	\end{picture}
    \begin{picture}(120,100)
		\put(0,0){\includegraphics[width=0.23\textwidth]{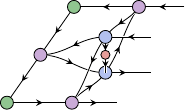}}
    \put(35,80){\textbf{b}} 
    \end{picture}
    \begin{picture}(100,100)
		\put(0,0){\includegraphics[width=0.20\textwidth,page=1]{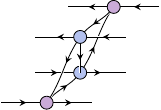}}
    \put(35,80){\textbf{c}} 
    \end{picture}
    \begin{picture}(100,100)
		\put(0,0){\includegraphics[width=0.23\textwidth]{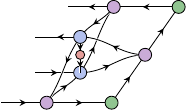}}
    \put(35,80){\textbf{d}} 
    \end{picture}
	\caption{a: Definition of the CTMRG environment tensors with the four corners (green) and edges (purple).
    b: Partial contraction of the left operator and infinite PEPS environment $L(O^{(1)})$.
    c: Definition of the horizontal transfer operator $T_H$, as built from the PEPS tensor and the environment tensors.
    d: Partial contraction of the right operator and infinite PEPS environment $R(O^{(2)})$.
    } \label{fig:ctmrgtensors}

\end{figure}

In other words, we extract the long-distance behavior of the two-point functions by studying the spectral decomposition of the horizontal transfer operator.
This will be dominated by the ratio of the dominant magnitude eigenvalues, and in particular by the gap in the spectrum.

Importantly, when trying to converge the CTMRG environments, for our PEPS it turns out that the algorithm is only stable for very specific values of the environment dimension $\chi$.
Inspecting the corner entanglement spectrum in Fig.~\ref{fig:corner_entanglement}, we can identify the cause: the (near-) degeneracies in the spectrum destabilize the iterations for any value of $\chi$ that cuts in between multiplets.
To remedy this issue, we employed a \emph{bootstrapping} procedure, where we find an initial stable value of $\chi$ for which the algorithm converges, and then inspect the spectrum after performing a single CTMRG iteration without truncation.
Then, we select a new value of $\chi$ according to the larger gaps in the spectrum, and repeat these steps to sequentially grow the entanglement in the environments.

\begin{figure}[t]
	\includegraphics[width=0.8\textwidth]{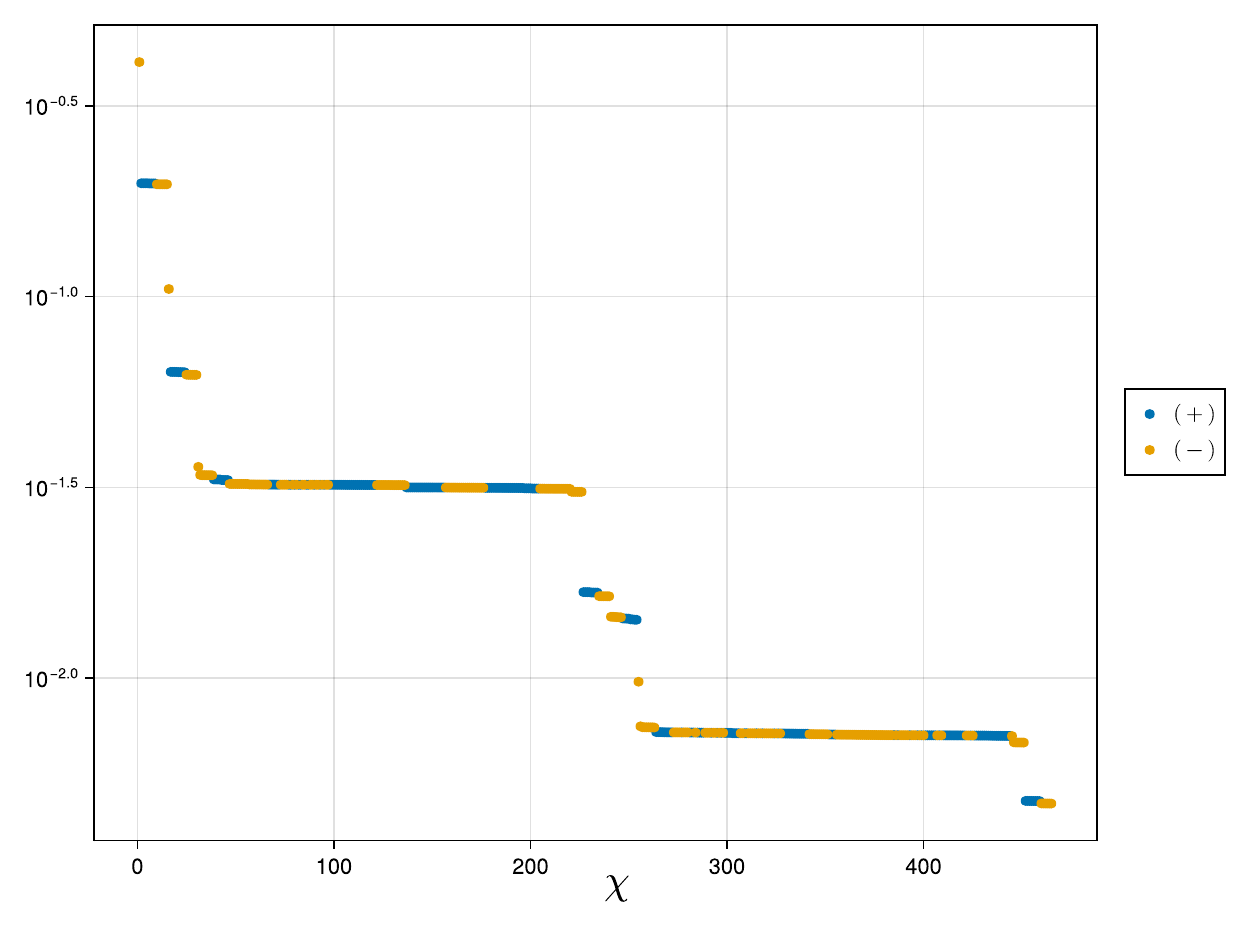}
	\caption{Corner entanglement spectrum for the CTMRG environment at boundary dimension $\chi = 232^{(+)} \oplus 233^{(-)} = 465$. The spectrum shows large clusters of near-degenerate values, with a few larger gaps in between. $+$ and $-$ refer to the even and odd subspace, respectively.}
	\label{fig:corner_entanglement}
\end{figure}

\end{widetext}
\end{appendix}

\end{document}